\documentstyle[12pt]{article}
\newcommand{\la}{\label}

\newcommand{\cL}{{\cal L}}

\newcommand{\be}{\begin{equation}}
\newcommand{\ee}{\end{equation}}
\newcommand{\ba}{\begin{eqnarray}}
\newcommand{\ea}{\end{eqnarray}}
\newcommand{\bastar}{\begin{eqnarray*}}
\newcommand{\eastar}{\end{eqnarray*}}

\begin{document}
\begin{titlepage}
\vskip 0.4truecm
\begin{center}
{ \bf \large \bf SHAFRANOV'S VIRIAL THEOREM AND \\ \vskip 0.3cm
                MAGNETIC PLASMA CONFINEMENT  \\
}
\end{center}

\vskip 1.0cm
\begin{center}
{\bf Ludvig Faddeev$^{\sharp \ddagger}$, Lisa Freyhult$^\star$, 
Antti J. Niemi$^{\star}$ and Peter Rajan$^\star$ } \\

\vskip 0.3cm

{\it $^\sharp$St.Petersburg Branch of Steklov Mathematical
Institute \\
Russian Academy  of Sciences, Fontanka 27 , St.Petersburg, 
Russia$^{\, \S}$ } \\
\vskip 0.3cm

{\it $^\star$Department of Theoretical Physics,
Uppsala University \\
P.O. Box 803, S-75108, Uppsala, Sweden }\\

\vskip 0.3cm

{\it $^{\ddagger}$Helsinki Institute of Physics \\
P.O. Box 9, FIN-00014 University of Helsinki, Finland} \\

\end{center}

\vskip 2.5cm

\rm
\noindent

Shafranov's virial theorem implies that nontrivial
magnetohydrodynamical equilibrium configurations must 
be supported by externally supplied currents. Here we extend
the virial theorem to field theory, where it relates to
Derrick's scaling argument on soliton stability. 
We then employ virial arguments to investigate a realistic 
field theory model of a two-component plasma, and  
conclude that stable localized solitons can exist
in the bulk of a finite density plasma. These solitons entail 
a nontrivial electric field which implies that purely 
magnetohydrodynamical arguments are insufficient for 
describing stable, nontrivial structures within the 
bulk of a plasma.

\noindent\vfill

\begin{flushleft}
\rule{5.1 in}{.007 in} \\
$^{\sharp}$ \small Supported by grants RFFR 99-01-00101 and INTAS 9606 \\
$^{\star}$ \small Supported by NFR Grant F-AA/FU 06821-308 \\
$^{\, \S}$ \small Permanent address
\\ \vskip 0.3cm
{\small  E-mail: \scriptsize
\bf FADDEEV@PDMI.RAS.RU, LISA.FREYHULT@TEORFYS.UU.SE, 
NIEMI@TEORFYS.UU.SE, PETER.RAJAN@TEORFYS.UU.SE}
\\
\end{flushleft}
\end{titlepage}

Ideal single fluid magnetohydrodynamics obeys an integral relation
which is known as Shafranov's virial theorem \cite{shaf},
\cite{frei}. It implies that an ideal 
magnetohydrodynamical system can not support nontrivial 
localized structures. Instead any nontrivial equilibrium 
configuration must be maintained by externally supplied currents,
a guiding principle in the design of contemporary magnetic
fusion devices.

Ideal magnetohydrodynamics is supposedly adequate for describing 
the ground state equilibrium geometry of a plasma. It also 
provides a starting point for a weak coupling Bolzmannian 
transport theory \cite{frei}. 
But as an effective mean field theory it lacks the kind of detailed
microscopic information which is needed to properly account 
for the electromagnetic interactions between the charged particles 
within the plasma. For this, ideal magnetohydrodynamics should be 
replaced by an appropriate classical field theory of charged
particles. With a firm microscopic basis
and established set of rules for systematic computations, a
field theory model can provide a rigorous basis for 
describing thermal fluctuations and dynamical
effects, including transport phenomena and issues 
related to plasma stability and confinement.

Here we extend Shafranov's virial theorem to
classical field theory where it yields a variant of Derrick's 
scaling argument, widely employed to inspect soliton stability. 
We then apply virial arguments to a realistic field theory model 
of plasma. In accordance with ideal 
magnetohydrodynamics, we conclude that the field theory does not 
support localized self-confined plasma configurations
in isolation, in an otherwise empty space. But when we 
inspect the finite density bulk properties of the field theoretical 
plasma, we find that the virial theorem does allow for the 
existence of stable solitons. These solitons describe extended 
collective excitations of charged particles in the otherwise 
uniform finite density environment. Our results are consistent with a 
recent proposal \cite{oma1}, that a finite density field theoretical plasma 
supports stable knotted solitons \cite{oma2}. Indeed, we 
expect that these solitons can be employed to describe a variety 
of observed phenomena. For example coronal loops that are 
present in solar photosphere are natural candidates.  The properties
of these solitons may also become attractive in fusion experiments, 
where their stability might help in the design of particularly 
stable magnetic geometries \cite{frei}.

Shafranov's virial theorem \cite{shaf} follows from the properties of 
the magnetohydrodynamical energy-momentum tensor $T^{\mu\nu}$ 
in the ideal single fluid approximation. Its spatial components 
are \cite{frei}
\be
T^{ij} \ = \ \rho v^i v^j \ + \ \left( p + \frac{1}{2} 
B^2 \right) \delta^{ij}
\ - \ B^i B^j
\la{emij}
\ee
while the purely temporal component coincides with the internal
energy density,
\be
T^{00} \ = \ \frac{1}{2} \rho v^2 \ + \ \frac{1}{2}
B^2 \ + \ \frac{p}{\gamma - 1}
\la{em00}
\ee
Here $\gamma$ is the ratio of specific heats. The fluid variables 
are the mass density $\rho$, the (bulk) fluid velocity $v^i$ and the 
pressure $p$, and $B^i$ is the magnetic field in 
natural units with $\mu_0 = 1$. The plasma evolves according to 
the Navier-Stokes equation which follows when we equate the 
divergence of the energy-momentum tensor with external 
dissipative forces.
These dissipative forces are present whenever the plasma is in motion, 
but cease when the plasma reaches a stable magnetostatic equilibrium 
configuration that minimizes the internal energy
\be
E \ = \ \int d^3 x \ T^{00}
\la{E}
\ee
Shafranov's virial theorem follows 
when we subject (\ref{E}) to a scale transformation 
of the spatial coordinates $x^i \to \lambda x^i$ with $\lambda$ 
a constant. For the magnetic field we select $B_i(x) \to \lambda^{2} 
B_i(\lambda x)$, as customary in Maxwell's theory. But 
for the pressure $p$ ideal magnetohydrodynamics does not 
supply enough information to
determine its behaviour under a scale transformation. 
For this we assume that the pressure is subject to the standard 
thermodynamic scaling relation of a thermally isolated gas, 
\be
p V^{\gamma} \ = \ constant
\la{eqst}
\ee
This implies that under a scaling $p(x) \to \lambda^{3\gamma} 
p (\lambda x) $. If we assume that the value $\lambda = 1$ corresponds
to an actual minimum energy configuration of the energy (\ref{E}), when viewed
as a function of $\lambda$ the energy (\ref{E}) then has an extremum 
at $\lambda = 1$. Consequently
\be
0 \ = \ \frac{\delta E(\lambda)}{\delta \lambda}_{|_{\lambda 
= 1}} \ = \ - \ 
\int d^3 x \left( 3 p + \frac{1}{2} B^2 \right) \ \equiv \ \ - \int 
d^3 x \ {T^{i}}_{i}
\la{shaf}
\ee
The magnetic contribution to the pressure is manifestly
positive definite. Furthermore, (collisionless) 
kinetic theory relates the 
pressure $p$ to the kinetic energy of the individual particles, 
which is similarly a positive definite quantity. The 
integrand in (\ref{shaf}) 
is then positive definite, and we conclude that under the present
assumptions non-trivial localized 
equilibrium configurations do not exist in ideal magnetohydrodynamics
\cite{shaf}, \cite{frei}. 

We have formulated our derivation of Shafranov's virial 
theorem so that it relates to Derrick's scaling argument 
in classical field theory \cite{derr}. For this, we consider a generic
three-dimensional Hamiltonian field theory model with classical 
action 
\be
S \ = \ \int dt d^3 x \ \cL(\psi) \ = \ \int dt d^3 x 
\left\{ \pi_\alpha {\dot \varphi}^\alpha \ - \ 
H[\pi,\varphi]\right\}
\la{act1}
\ee
The fields $\psi_\alpha \sim (\pi_\alpha , 
\varphi^\alpha)$ are canonical conjugates with Poisson bracket
\be
\{ \pi_\alpha(x) , \varphi^\beta (y) \} \ = 
{\delta_\alpha}^\beta (x - y)
\la{pb}
\ee
Notice that the time derivative in (\ref{act1}) acts asymmetrically.
But in the sequel it will be useful to consider symmetrized
quantities, and for this we generalize the time derivative 
term by a canonical transformation into
\be
\int dt d^3 x \  \pi_\alpha {\dot \varphi}^\alpha \ \to \
\int dt d^3 x \ \left[ \ a {\dot \pi_\alpha}\varphi^\alpha \ + \ (1-a) 
\pi_\alpha {\dot \varphi}^\alpha \ \right]
\la{symmcan}
\ee
where $a$ parametrizes the canonical transformation. 

We assume that the Hamiltonian $H$ is a functional of the 
fields $\psi = (\pi,\varphi)$ and their first derivatives 
only, with no explicit 
dependence on the space coordinates $x^i$ and time $t \equiv x^0$.
We then obtain the energy-momentum tensor directly from Noether's 
theorem: Since there is no explicit dependence on $x^\mu$
\be
\frac {\partial \cL}{\partial x^\mu} \ = \ \frac{\delta \cL }{\delta 
\psi_\alpha} \partial_\mu \psi_\alpha \ + \ \frac{\delta \cL}{\delta 
\partial_\nu \psi_\alpha} \partial_\mu \partial_\nu \psi_\alpha
\la{emt1}
\ee
and by employing the equations of motion we identify the components of
the energy-momentum tensor with the ensuing four conserved currents
\be
{T^\mu}_\nu \ = \ \frac{\delta \cL} {\delta \partial_\mu \psi_\alpha}
\partial_\nu \psi_\alpha \ - \ {\delta^\mu}_{\nu} \cL
\la{emt2}
\ee
In general (\ref{emt2}) fails to be symmetric. But in the following
we find it useful to consider symmetrized quantities,
and for this we can re-define
\be
{T^\mu}_\nu  \ \to \ {T^\mu}_\nu \ + \ \partial_\rho
{X^{\rho \mu}}_\nu 
\la{emt3}
\ee
where ${X^{\rho \mu}}_\nu = -{X^{\mu \rho}}_\nu$ has no effect on the 
dynamics. (Note that if the theory fails to
be Lorentz invariant, in general there will be no symmetry 
between the momentum flux ${T^0}_i$ and the energy flux ${T^i}_0$.)

We are interested in a scale transformation of the 
spatial coordinates $x^i \to \lambda x^i$, which sends 
$\psi_\alpha(x) \to \lambda^{D_\alpha} \psi_\alpha (\lambda x)$. 
Here $D_\alpha$ is the scale dimension of the field $\psi_\alpha$.
By considering an infinitesimal transformation with
$\lambda = 1 + \epsilon$ we find for the generator $\delta_S$ 
of the scale transformation
\ba
\delta_S \pi_\alpha  \ & = & \  x^i \partial_i \pi_\alpha  \ 
+ \ D_\pi^\alpha \pi_\alpha
\la{scp} \\
\delta_S \varphi^\alpha  \ & = & x^i
\partial_i \varphi^\alpha \ + \ D^\varphi_\alpha \varphi^\alpha
\la{scq}
\ea
For the energy density this yields
\be
\delta_S {T^0}_0  \ = \ \left\{ \ - {T^i}_{i} \
+ \ \partial_i (x^i  {T^0}_0 ) \ + \
\sum_{\alpha} D_\alpha \left[ \frac{ \delta  {T^0}_0}
{ \delta \psi_\alpha} \psi_\alpha \ + \  \frac{ \delta  {T^0}_0}
{ \delta \partial_k \psi_\alpha} \partial_k 
\psi_\alpha \right] \ \right\}
\la{Dact2}
\ee 
In general the scale dimensions can be arbitrary, and there
is no {\it a priori} relation between the different 
$D_\alpha$.  But if the scale transformation is 
a canonical transformation it must preserve Poisson brackets,
which implies that the scale dimensions of a canonical
pair are subject to
\be
D_\pi^\alpha \ + \ D^\varphi_\alpha \ = \ 3
\la{DD}
\ee 
The generator $\delta^C_S$ of such a canonical scale
transformation can be computed from Noether's theorem,
and by properly selecting the value of $a$
in (\ref{symmcan}) we arrive at the symmetrized form
\be
\delta^C_S \ = \ \int d^3 x \ x^i {T^0}_i
\la{S}
\ee

We are interested in a refinement of ideal magnetohydrodynamics,
a microscopic field theory model of a two-component plasma 
with negatively charged electrons $(e)$ and positively 
charged ions $(i)$ and classical (first-order) 
Lagrangian \cite{oma1},
\[
\cL \ = \ E_k \partial_t A_k  \ + \  
\frac{i}{2} \left( \psi_e^* \partial_t \psi_e \ - \ \partial_t
\psi_e^* \psi_e \ + \ \psi_i^* \partial_t \psi_i \ - \ 
\partial_t \psi_i^* \psi_i \right) \ - \ \frac{1}{2} E_k^2 - \ 
\frac{1}{2}B_k^2
\]
\be 
 - \ \frac{1}{2m} | (\partial_k +
i e A_k) \psi_e |^2  \ - \ \frac{1}{2M} | (\partial_k -
i e A_k) \psi_i |^2 \ + \ A_0 ( \ \partial_k E_k \ - \ e {\psi_e}^*
\psi_e \ + \ e {\psi_i}^* \psi_i \ )
\la{act} 
\ee
Here $\psi_e$ and $\psi_i$ are (complex) non-relativistic 
Hartree-type fields 
that describe electrons and ions with masses $m$ and $M$ and electric
charges $\pm e$ respectively, together with their electromagnetic
interactions. Note that we have realized Maxwell's theory canonically so that
the electric field $E_i$ and spatial gauge field $A_i$ form a canonical
pair, with the temporal $A_0$ a Lagrange multiplier that enforces Gauss' law. 
Since the time derivative appears linearly in the charged fields, 
the action (\ref{act}) admits a proper Hamiltonian
interpretation with $\psi_{e,i}^*$ the canonical conjugates of 
$\psi_{e,i}$. Notice that for definiteness we 
have chosen both charged fields to be 
commuting. This should be adequate in the Bolzmannian limit, relevant in 
conventional plasma scenarios where the temperature is sufficiently 
high so that bound states (hydrogen atom) are prevented but not 
high enough for relativistic corrections to become important. Notice
that we have also introduced an appropriate symmetrization of
the form (\ref{symmcan})
in the time derivative terms of the charged fields. 
Finally, besides the terms that we have displayed in (\ref{act}) 
we implicitely assume the presence of chemical potential
terms that ensure overall charge neutrality. However, fr 
the present purposes such terms are redundant and will either
remain implicit, or will be enforced by appropriate boundary conditions.

We propose that the advantage of (\ref{act}) over ideal 
magnetohydrodynamics is, that (\ref{act}) provides a 
firm microscopic basis for systematically
computing various properties of a plasma. For example an 
appropriate version of the equation of state (\ref{eqst}) can be derived
from (\ref{act}). In particular,
(\ref{act}) yields immediately the standard 
electromagnetic many-body Schr\"odinger equation for a 
gas of electrons and ions. 

The energy-momentum tensor ${T^{\mu}}_{\nu}$ 
can be computed directly from (\ref{emt2}). After we introduce an 
appropriate symmetrization which ensures manifest gauge
invariance, we find for the energy density 
\be
{T^{0}}_{0} = \ 
\frac{1}{2\mu} \{ \sin^2 \alpha | D_k \psi_e |^2 \ + \ \cos^2 
\alpha | D^*_k \psi_i |^2 \} \ + \ \frac{E^2}{2} \ + \
\frac{B^2}{2} \ - \ A_0 \left( \partial_i E_i + 
e [ \psi^{*}_i \psi_i -  \psi^{*}_e \psi_e ]\right)
\la{ene}
\ee
where $D_k = \partial_k + i e A_k$ and 
$\mu = m \sin^2 \alpha = M \cos^2 \alpha$ is the 
reduced mass. For the spatial components of the energy-momentum 
tensor we find similarly, with the help of the equations of motion
\[
{T^{i}}_{k} \ = \ E_i E_k \ + \ B_i B_k \ - \ \frac{1}{2\mu}
\biggl\{ \sin^2 \alpha \bigl[ (D_i \psi_e)^* ( D_k \psi_e) \ + \
(D_k \psi_e)^* ( D_i \psi_e)  \bigr] 
\] 
\be
+ \ \cos^2 \alpha \bigl[ (D^*_i \psi_i)^* ( D^*_k \psi_i) \ + \
( D^*_k \psi_i)^* ( D^*_k \psi_i) \bigr] \biggr\}
\ - \ {\delta^i}_k \ \cL 
\la{tij}
\ee
Finally, for the generator of the canonical 
scale transformation we get
\be
\delta^C_S \ = \int d^3 x \ x^k {T^0}_k =  \int d^3 x \ x^k 
\left[ E_i F_{ki} + \frac{i}{2} \{ \psi_e^* D_k \psi_e  -  
D^*_k \psi_e^* \psi_e  +  \psi_i^* D^*_k \psi_i  -  
D_k \psi_i^* \psi_i \} \right]
\la{canD}
\ee 
It yields the following gauge covariantized version of (\ref{scp}),
(\ref{scq}),
\ba
\delta^C_S E_k \ & = & \ x^i \partial_i E_k \ + \ 2 E_k \ + \ x^k( 
\partial_i E_i \ + \
e [ \psi^{*}_i \psi_i -  \psi^{*}_e \psi_e ]) \\
\delta^C_S A_k \ & = & \ x^i \partial_i A_k \ + \ A_k - 
\partial_k ( x^i A_i ) \\
\delta^C_S \psi_{e,i} \ & = & \ x^i \partial_i 
\psi_{e,i} \ + \frac{3}{2} \psi_{e,i}
\ \pm \ i e x^i A_i \psi_{e,i} 
x^i \partial_i \psi^*_{e,i} + \frac{3}{2}\psi^*_{e,i} \
\la{Dact2b}
\ea
In particular, for each of the canonical variable
$(\psi_{e,i} , \psi^*_{e,i})$ the scale dimension is $3/2$
so that the canonical scale generator commutes with
the number operators for the charged particles
\be
\{ \delta^C_S \, , \, N_{e,i} \} \ = \ \delta^C_S \int d^3x \
\psi^*_{e,i} \psi_{e,i} \ = \ 0
\la{commcan}
\ee

We now proceed to inspect the consequences of Shafranov's 
virial arguments. For this we remind that a static minimum
energy configuration must be a stationary point of the
energy (\ref{E}), (\ref{ene}) under any {\it local} variation of the
fields. Since the scale transformation (\ref{scp}), (\ref{scq})
is a non-local variation it  
does not need to leave the energy intact, unless 
it also preserves the pertinent boundary conditions. 
To determine these boundary conditions, we consider the 
plasma in two different physical environments:
 
In the first scenario we have an isolated, 
localized plasma configuration in an otherwise 
empty space, with a definite number of charged particles
\be
N_{e} + N_i \ = \ \int d^3x \ \left( \ \psi_e^* \psi_e \ + \
\psi_i^* \psi_i \ \right)
\la{first}
\ee
Since the canonical scale generator commutes with the individual
number operators (\ref{commcan}), the ensuing variation of the fields
is consistent with the boundary condition that the number of 
particles remains intact. By a direct computation we 
then find for a static stationary point of the energy, 
\[
0 \ = \ \delta^C_S \int d^3 x \ {T^0}_0 \ = \ - \ \int d^3 x \ {T^i}_i 
\]
\be
= \ - \ \int d^3 x \left( \ - \frac{1}{\mu} \{ \sin^2 \alpha 
| D_k \psi_e |^2 \ + \ \cos^2 
\alpha | D^*_k \psi_i |^2 \} \ - \ \frac{E^2}{2} \ - \
\frac{B^2}{2} \right)
\la{varinr}
\ee
Since the trace of the spatial stress tensor is a 
sum of positive definite terms, in analogy with Shafranov's 
virial theorem in ideal magnetohydrodynamics (\ref{shaf}) we 
conclude that there can not be any nontrivial stationary 
points. This means that in an otherwise 
empty space an initially localized plasma configuration
can not be confined solely by its internal electromagnetic 
interactions. additional interactions such as gravity must be
present. Otherwise the canonical scale transformation
dilutes the plasma by expanding its volume while 
keeping the number of the charged particles intact,
until the collective behaviour of the
plasma becomes replaced by an individual-particle behaviour
of the charged constituents. 

The second physical scenario of interest to us
describes the bulk properties of a plasma: 
We are interested in an initially localized
plasma configuration, located
within the bulk of a finite density plasma background. 
In this case the relevant boundary condition on the 
charged fields states, that at large distances their 
densities approach a non-vanishing constant value 
$\rho_0$ which is the density of the uniform
background plasma,
\be
|\psi_{e,i} \, |^2 \ \stackrel{r\to\infty}
{\longrightarrow} \ \rho^2_0 
\la{second}
\ee 
The canonical scale transformation assigns a non-trivial
scale dimension to the charged fields. Consequently it 
can not leave the asymptotic particle density intact, 
and fails to be consistent with the boundary 
condition (\ref{second}) unless $\rho_0 = 0$.
Instead of the canonical version of the scale transformation,
we need to employ a non-canonical version of 
(\ref{scp}), (\ref{scq}) where the scale dimensions of 
the charged fields vanish, $D_\psi = 
0$. When we perform the ensuing variation of the fields
in the energy density (\ref{ene}), instead 
of (\ref{varinr}) we find
\be
\delta_S \int d^3 x \ {T^0}_0 \ = \
\int d^3 x \ \left[ \frac{E^2}{2} \ + \ \frac{B^2}{2} 
\ - \ \frac{1}{2\mu}\left( \sin^2\alpha | D_k \psi_e |^2 \ + \ \cos^2 \alpha
|D^*_k \psi_i |^2 \right) \right]
\la{relvar}
\ee
Now the integrand acquires both positive and negative contributions,
which implies that a virial argument can not exclude the
existence of stable finite energy solitons. Indeed, 
in \cite{oma1} it has been argued that
stable knotted solitons are present.
These solitons are formed within the bulk of the plasma,
in an environment with an asymptotically constant background density. 
A physical example of such an environment is the solar 
photosphere, the solitons are natural candidates 
for describing stable coronal loops. Another, somewhat more 
hypothetical example could be the ball lightning, in the background
of Earth's atmosphere. Such solitons could also become 
relevant in identifying particularly
stable plasma configurations in fusion experiments, when the plasma
is kept at finite density by the boundaries of an appropriate vessel.

We shall now proceed to demonstrate, that the virial 
theorem (\ref{relvar}) is also consistent with an 
appropriate canonical scale tranformation. For this 
we first notice that excluding the kinetic terms, the 
Lagrangian (\ref{act}) coincides with that of 
relativistic scalar electrodynamics with two flavors 
of scalar fields,
\be
\cL \ = \  |(\partial_\mu + i A_\mu) \phi_1|^2 \ + \ 
|(\partial_\mu - i A_\mu) \phi_2|^2  \ - \ V(\phi) \ - \
\frac{1}{4} F_{\mu\nu}^2
\la{sqed}
\ee
Here we have included a Higgs potential $V(\phi)$, 
to ensure a non-vanishing asymptotic value for the charged fields. 
For example, we can choose
$V(\phi) \propto (\phi_1^2 + \phi_2^2 - \rho_0^2)^2$. 
The Hamiltonian version of (\ref{sqed}) is 
\[
\cL \ = \ \pi_1^* \partial_0 \phi_1 \ + \ \pi_1 \partial_0 \phi_1^* \ + \
\pi_2^* \partial_0 \phi_2 \ + \ \pi_2 \partial_0 \phi_2^* \ + \
E_i \partial_0 A_i \ - \ |(\partial_k + iA_k)\phi_1|^2 - 
|(\partial_k - iA_k)\phi_2|^2 \]
\be
- \pi^*_1 \pi_ 1 \ - \ \pi^*_2 \pi_2 \ - \ V(\phi)
\ - \ \frac{E^2}{2} \ - \ \frac{B^2}{2}
- A_0 ( \partial_i E_i \ + \ i \pi^*_1 \phi_1 \ - \ i \pi_1 \phi^*_1
\ - \ i \pi^*_2 \phi_2 \ + \ i \pi_2 \phi^*_2 )
\la{hsqed}
\ee
Notice that now the charged fields are canonically 
independent variables,
a consequence of Lorentz invariance.
The energy-momentum tensor can be computed directly from
(\ref{emt2}). With a proper symmetrization it becomes
fully symmetric, as it should since the theory is Lorentz invariant. 
For the energy density we find
\[
{T^0}_0 \ = | D_k \phi_1|^2 \ + \ | D^*_k \phi_2 |^2 \ + \
\frac{E^2}{2} \ + \ \frac{B^2}{2} \ + \ \pi^*_1 \pi_1 \ + \ \pi^*_2 \pi_2
\ + \ V(\phi) \]
\be
- \ A_0 \{ \partial_i E_i \ + \ i (\pi^*_1 \phi_1 - \pi_1 \phi^*_1)
\ - \ i (\pi^*_2 \phi_2 - \pi_2 \phi^*_2)  \}
\la{relene}
\ee
The momentum flux is
\be
{T^0}_k \ = \ E_i F_{ki} \ + \ \pi_1^* D_k \phi_1 \ + \
\pi_1 D^*_k \phi_1^* \ + \ \pi_2^* D^*_k \phi_2 \ + \
\pi_2 D_k \phi_2^ *
\la{relmom}
\ee 
so that instead of (\ref{Dact2b}), we find that the canonical
scale dimensions of the charged scalar fields now 
vanish. As a consequence the canonical scale 
transformation is consistent with the relevant 
boundary condition that in the $r\to \infty$ limit 
the system approaches a Higgs vacuum
$\phi_1^2 + \phi_2^2 \to \rho_0^2 $. This means that
the canonical scale transformation must now leave the
energy of a static stationary point intact which leads to
the following virial theorem
\[
0 \ = \ \delta_S^C \int d^3 x \ {T^0}_0 \ = \ - \int d^3 x \ {T^i}_i
\]
\be
= \int d^3 x \ \left[ \frac{E^2}{2} \ + \ \frac{B^2}{2} 
+ 3(\pi_1^* \pi_1 + \pi_2^* \pi_2)
\ - \ | D_k \phi_1 |^2 \ - \ |D^*_k \phi_2 |^2 \right]
\la{relvar2}
\ee
Since the contribution to the pressure 
from the charged fields is negative, the virial theorem 
can not exclude stable static solitons.

Finally, we note that even though the static 
sectors of the two theories (\ref{act}) and (\ref{sqed}) 
are very similar, these theories actually have a quite
different physical content: In the relativistic 
case we may consistently set $\pi_1 = \pi_2 = E_i = A_0 =0$ in the static 
equations of motion. This reduces the energy density to a 
functional form which is manifestly 
magnetohydrodynamical (\ref{em00}),
\be
E \ = \ \int d^3x \ \left[ | D_k \phi_1|^2 \ + \ | 
D^*_k \phi_2 |^2 \ + \ V(\phi) \ + \
\frac{B^2}{2} \right]
\la{sh1}
\ee
But since the canonical scaling dimensions of 
the charged fields now vanish, the virial
theorem does not exclude the existence of purely magnetic
solitons. On the other hand, if in the non-relativistic 
case (\ref{act}) we set the electric field to vanish, the 
equations of motion become inconsistent unless the 
electron and ion charge densities are everywhere identical. 
This leads to a contradiction whenever the Hopf invariant 
is nontrivial \cite{oma1}, \cite{oma2}. 
Hence solitons with a nontrivial Hopf
invariant are necessarily accompanied with a nontrivial 
electric field. In particular, this means that their properties 
can not be consistently inspected by pure magnetohydrodynamics.

\vskip 0.3cm
In conclusion, we have extended Shafranov's virial theorem from ideal 
magnetohydrodynamics to classical field theory and related
it with Derrick's scaling argument. We then employed the
virial theorem to inspect soliton stability in a realistic field
theory model of a two component plasma. In line with 
ideal magnetohydrodynamics, a scaling argument reveals that 
the field theory model does not support stable
isolated solitons in an otherwise empty space. 
But the virial theorem does allow for the existence 
of stable solitons within the bulk of the plasma.
These solitons are accompanied by a nontrivial electric field, hence
they can not be probed by magnetohydrodynamics alone. 
We suggest that these solitons are relevant in describing 
coronal loops in solar photosphere, maybe even ball lightning
in Earth's atmosphere. They might also become useful in 
the design of particularly stable magnetic fusion geometries.

\vskip 0.5cm
We thank A. Bondeson, R. Jackiw, S. Nasir, A. Polychronakos
and G. Semenoff for discussions.

\end{document}